# Interpretable Artificial Intelligence for Detecting Acute Heart Failure on Acute Chest CT Scans

Silas Nyboe Ørting † (PhD) (1), Kristina Miger † (MD) (2)(3), Anne Sophie Overgaard Olesen (MD) (2)(3), Mikael Ploug Boesen (MD, PhD, Professor) (3)(4), Michael Brun Andersen (MD, PhD, Associate Professor) (5), Jens Petersen (PhD, Associate Professor) (1), Olav W. Nielsen ‡ (MD, PhD, DMSc, Professor) (3), Marleen de Bruijne ‡ (PhD, Professor) (1)(6)

1) Department of Computer Science, University of Copenhagen, Denmark
2) Department of Cardiology, Copenhagen University Hospital – Bispebjerg and Frederiksberg, Denmark
3) Department of Clinical Medicine, University of Copenhagen, Denmark
4) Department of Radiology, Copenhagen University Hospital – Bispebjerg and Frederiksberg, Denmark
5) Department of Radiology, Copenhagen University Hospital – Herlev Gentofte, Denmark
6) Department of Radiology and Nuclear Medicine, Erasmus MC – University Medical Center Rotterdam, Rotterdam, The Netherlands

†: These authors contributed equally as first authors.

‡: These authors contributed equally as last authors.

**Summary statement:**

Subjects at risk of acute heart failure can be detected using an explainable tree-based model trained on cardiac and pulmonary structures segmented from acute thoracic CT scans.

**Key points:**

1) We developed an explainable artificial intelligence model for detecting radiological signs of acute heart failure on acute chest CT scans.

2) The model demonstrated good discriminatory value, comparable to clinical radiologists.

3) The combination of key features with a tree-based model and visual segmentations allows for transparent, step-by-step predictions that can easily be interpreted by physicians.

**Abstract**

**Introduction:** Chest CT scans are increasingly used in dyspneic patients where acute heart failure (AHF) is a key differential diagnosis. Interpretation remains challenging and radiology reports are frequently delayed due to a radiologist shortage, although flagging such information for emergency physicians would have therapeutic implication. Artificial intelligence (AI) can be a complementary tool to enhance the diagnostic precision.

We aim to develop an explainable AI model to detect radiological signs of AHF in chest CT with an accuracy comparable to thoracic radiologists.

**Methods:** A single-center, retrospective study during 2016-2021 at Copenhagen University Hospital – Bispebjerg and Frederiksberg, Denmark. A Boosted Trees model was trained to predict AHF based on measurements of segmented cardiac and pulmonary structures from acute thoracic CT scans. Diagnostic labels for training and testing were extracted from radiology reports. Structures were segmented with TotalSegmentator. Shapley Additive explanations values were used to explain the impact of each measurement on the final prediction.

**Results:** Of the 4,672 subjects, 49% were female. The final model incorporated twelve key features of AHF and achieved an area under the ROC of 0.87 on the independent test set. Expert radiologist review of model misclassifications found that 24 out of 64 (38%) false positives and 24 out of 61 (39%) false negatives were actually correct model predictions, with the errors originating from inaccuracies in the initial radiology reports.

**Conclusion**: We developed an explainable AI model with strong discriminatory performance, comparable to thoracic radiologists. The AI model's stepwise, transparent predictions may support decision-making.

# 1. Introduction:

Acute heart failure (AHF) is a severe health condition with high mortality and rehospitalization rates. It is a major cause of emergency hospital admissions worldwide, and a leading cause of hospitalizations in elderly patients. Early and accurate diagnosis is crucial for initiating timely and effective treatment, improving patient outcomes(1).

Chest computed tomography (CT) scans are increasingly used as a diagnostic tool in early triage for patients with dyspnea – a key symptom in AHF (2). CT imaging is particularly valuable in patients with co-existing pulmonary conditions, for detecting pulmonary embolism, and in supporting differential diagnosis. However, the systematic assessment of chest CT scans for radiological signs of AHF remains challenging, especially without a universally accepted standard for interpretation among radiologists.

As CT utilization increases, the volume of CT images requiring interpretation will outpace the capacity of available radiologists, leading to diagnostic delays. This is further aggravated by the shortage of radiologists worldwide. This backlog may subsequently delay timely communication of radiological findings to the treating physician and therefore delay appropriate initiation of treatment for AHF. Consequently, there is a pressing need for decision-support tools capable of assisting radiologists in identifying patients with features of AHF, while concurrently alerting emergency physicians to the likelihood of AHF prior to the availability of the formal radiology report.

Artificial intelligence has emerged as a complementary tool that enhances diagnostic precision in cardiovascular medicine, diminishing the chest CT reading time for cardiothoracic radiologists by 22%(3). However, the application of AI in the context of AHF is limited. To our knowledge, no study has yet evaluated the performance of an AI model in detecting radiological signs of AHF on acute chest CT scans.

We hypothesize that an explainable AI model can be developed to detect radiological signs of AHF on chest CT with accuracy comparable to that of thoracic radiologists. We aim to support clinicians by flagging potential patients with AHF during triage, directing radiologists' attention to relevant findings, and enabling timely cardiology follow-up.

**2. Materials and Methods**

We trained a Boosted Trees(4) model to predict AHF based on measurements of cardiac and pulmonary structures.

2.1 Study Design

A single-center, retrospective study using CT scans to develop an AI model for detecting AHF (Figure 1).

2.2 Data

We included patients with at least one acute thoracic CT scan and subsequent radiology report at Copenhagen University Hospital – Bispebjerg and Frederiksberg, Denmark, during 2016-2021.

We excluded patients that had been included in a separate prospective observational study in the same period (the FACTUAL study) to allow for validation(5): incomplete reports; partial lung scans or scans that failed to be read.

2.3 Ground Truth

Ground truth of radiological signs of AHF were extracted from the radiology reports: pulmonary congestion, pulmonary oedema, heart failure, and decompensation (Supplementary Table S1). We manually reviewed all reports mentioning any of these AHF indicators and corrected any labeling errors. The labels were reviewed for technical correctness by SNØ. Any unclear case was additionally reviewed by KCM, ASOO and OWN. This approach was chosen to maximize the dataset and reflect real-world diagnostic reporting by radiologists.

2.5 Data Partitions

The data was randomly split into training and test data on subject level (Table 1).

Parameter tuning and feature selection were performed on the training data using cross validation, with folds stratified by subjects and kept consistent throughout each step. For test subjects with multiple studies at different time points, only the latest study was included. Multiple CT acquisitions and reconstructions were predicted separately and averaged to a single study-level prediction.

2.6 Segmenting Structures

TotalSegmentator Version 1 (6) was used to segment left/right atrium, left/right ventricle, myocardium, pulmonary artery, vena cava inferior, lung, pleural effusion, and pericardial effusion.

The total lung was defined as the union of lung and pleural effusion; lung tissue as lung without pleural effusion; lung boundary as a 10 mm band next to the boundary; and total heart as the union of left/right atrium, left/right ventricle, and myocardium(6) (Supplementary Figure S1).

2.7 Features

We selected radiological features from a prior study on detecting AHF via chest CT (5) and parameters physiologically associated with heart failure were also added.

We measured: the volume of total lung, pleural effusion, total heart, left/right ventricle, left/right atrium, and pericardial effusion; the diameter of vena cava inferior and the pulmonary artery; the density of lung tissue, lung boundary, vena cava inferior, and right atrium.

We derived volume ratios for total heart and pleural effusion (to total lung) and for left/right atrium, left/right ventricle, and pericardial effusion (to total heart). We calculated the modified Z score for all volumes. A contrast-only version was derived for each density measurement.

Missing feature values due to segmentation errors were handled by including them as possible splits in the decision trees.

2.8 Training the Prediction Model

We used a three-step approach to optimize the Boosted Trees model(4). In step one, the model parameters were tuned using all features as input. In step two, the tuned model was used to select the most important features. In step three, the model parameters were re-tuned using the selected features. We used 3-fold cross validation and area under the ROC curve (AUROC) to measure performance on the training set.

Parameter tuning was done using grid search (Supplementary Table S2). Initial feature selection was done using forward feature selection based on an XGBoost internal measure of feature importance. The selected features and parameters were used to train a model that was analyzed with Shapley addi-tive explanation (SHAP) values (7). We then manually removed features deemed to be clinically unimportant and redundant, and trained a final model that was evaluated on the test data.

2.9 Evaluation

The segmentations were evaluated qualitatively by visual inspection. All lung and pulmonary artery segmentations were inspected via projections along the three axes (Supplementary Figure S2). A set of CT volumes were selected based on extreme feature values (N=9; smallest/largest lung, heart, pulmonary artery, VCI; largest pleuraeffusion), and inspected using 3D Slicer (Supplementary Figure S3).

Model predictions were evaluated internally using a ROC curve, AUROC and a confusion matrix of thresholded predictions. The threshold was defined as the cutoff leading to 5% false positive predictions on the training data. All studies in the test set with an incorrect prediction under this threshold were reviewed by MBA (MD, PhD, Associate Professor) providing a full radiological description of heart and lung findings, and a 5-point score for the probability of AHF.

We used SHAP beeswarm and bar plots to explain the general effect of each feature in the trained prediction model and SHAP waterfall plots to explain the effect of each feature on individual predictions.

Evaluation of the model's performance on external data was not done in this study but will be the focus of future studies.

**3. Results**

3.1 Data

Inclusion and exclusion are summarized in Figure 2. We extracted radiology reports covering 6101 studies of 5479 subjects. We excluded 157 subjects due to incomplete reports, and 234 subjects due to being included in the FACTUAL study. In total, we included 5678 studies of 5140 subjects.

For the CT scans, we extracted 16330 scans covering 5876 studies of 5294 subjects. Of these, 352 subjects were excluded because there was no corresponding report, 116 subjects were excluded due to errors in reading the CT scans, and 154 subjects were excluded because the scan did not include the full lung field. In total, we included 12348 CT scans - comprising i.e., pre- and post-contrast and various reconstructions - covering 5111 studies of 4672 subjects (Supplementary Table S3). The 4672 subjects were split in training (N=3148) and test (N=1524) groups.

3.2 Obtaining ground truth from radiology reports

We extracted 295 positive and 278 negative "congestion" findings; 34 positive and 0 negative "edema" findings; 28 positive and 7 negative "heart failure" findings; and 153 positive and 11 negative "decompensation" findings. After review of the reports, these numbers were corrected to 281 positive and 275 negative "congestion" findings; 34 positive and 0 negative "edema" findings; 30 positive and 5 negative "heart failure" findings; and 153 positive and 12 negative "decompensation" findings.

In total after correction, there were 498 positive findings and 292 negative findings. Reports without any positive or negative findings were assumed negative. This resulted in 358 subjects labeled as positive for AHF and 4314 labeled as negative.

3.3 Visual Inspection of Segmentations

Approximately 80% of the pulmonary artery segmentations were deemed to be of reasonable quality. All lung segmentations were deemed to be of reasonable quality. The visual inspection of extreme cases (smallest/largest segmented size) found the quality of segmentations to be good, except for the smallest heart and VCI, where the segmentation failed in both cases. Inspection of five more small VCI segmentations indicated large variation in segmentation quality, whereas inspection of five more small heart segmentations indicated reasonable quality. Another potential issue was in the boundaries between substructures in the heart, which were not visible in non-contrast scans. Despite this, TotalSegmentator still provided a plausible segmentation of the structures.

3.4 Training the Prediction Model

The initial and final parameter tuning resulted in the parameters listed in Supplementary Table S2, and these were used to train a model on the full training data.

The final set of selected features were evaluated by analyzing a SHAP Beeswarm plot. This clearly indicated that mean lung density and heart volume provided almost no discriminating information, as both features were redundant when median lung density and absolute heart Z score were included (Supplementary Figure S4). We therefore removed mean lung density and heart volume as features. A SHAP Beeswarm plot of the final model trained without mean lung density and heart volume, is included in Figure 3 and is visually identical to Figure S4, except for the excluded features.

As a result, the final model included twelve features: mean lung boundary (HU), pleural ratio (pleural effusion volume/total lung volume), left atrial volume, pleural effusion volume, median lung density (HU), mean density of vena cava inferior (HU), right atrial volume, right ventricle ratio (right ventri-cle/total heart volume), absolute heart Z score, age, right atrial density (HU), and the diameter of vena cava inferior.

3.5 Evaluation of the Model Performance

Predictions on the train data yielded an AUROC of 0.93, and on the test data an AUROC of 0.87 (Figure 4). Classification of test studies into either AHF or non-AHF was performed with the threshold (t=0.26840377), which yielded a 5% false positive rate on the training data.

The classification resulted in 126 errors (8.3 %), consisting of 64 false positive cases and 62 false negative cases (Table 2). We found similar error rates for females and males, though there may be a trend towards higher false positive rate in males (0.06 vs 0.03 in females) and a higher false negative rate in females (0.53 vs 0.46 in males).

A thoracic radiologist (MBA) then conducted a review of these 126 errors to assess the validity of the model output, and to characterize the nature of the errors (Table 3). Upon error review, the radiologist identified discrepancies with the labels extracted from the clinical radiology reports in 48 (38%) of cases and disagreed with the model classifications in 72 (58%) of cases. The review identified 24 of 64 false positives as true positives, and 24 of 61 false negatives as true negatives. One false negative case was not reviewed due to issues with PACS. Prediction and review of two false positive cases are illustrated in Figure 5.

**4. Discussion**

4.1 Summary

We developed and internally validated an explainable AI model to detect radiological signs of AHF on emergency chest CT scans. The model performed well, achieving an AUROC of 0.87—comparable to assessments by thoracic radiologists. It relies on a small set of features, all directly related to the pathophysiology of AHF, and uses a tree-based structure that makes its decision-making process transparent and easy to follow. Clinicians can visually review these features through image segmentations, helping them understand, trust, and explain the model's predictions.

4.2 Previous literature

Although there is a growing body of research of AI in cardiovascular disease, no existing studies have developed AI algorithms specifically for identifying radiologic signs of AHF on CT scans in undifferentiated patients (8). Few studies have explored AI for pulmonary congestion, but those focus on other clinical contexts(8–10)(9). Our study adds insights into how an AI model can detect radiological signs of AHF in a general emergency population.

4.3 Ground Truth Definition and Model Evaluation

The error analysis indicates that the quality of labels from radiology reports may be a concern.

Although real-time radiologist interpretations can act as ground truth, these labels vary in quality, as there currently is no standardized protocol for reporting AHF signs on CT, leading to underreporting and label variability. Despite limitations, using immediate reports allowed for a large cohort, though external validation with standardized datasets remains essential(11).

The error analysis found that 38% of the prediction errors where attributable to label quality, indicating that performance could improve with more accurate ground truth and a structured reporting approach during the development. Nonetheless, the relatively high false negative rate remains a concern. While it could be mitigated by lowering the decision threshold, doing so would inevitably increase the number of false positives.

The cost of developing our AI models is greatly reduced by relying on off-the-shelf segmentation software. The downside is that this software may not be optimized for the clinical setting, which could lead to reduced quality of predictions. Our inspection of lung segmentations indicated good quality but a decrease in segmentation quality should be expected for finer structures, as observed for the pulmonary artery, possibly explaining its exclusion in the final feature selection. TotalSegmentator produced plausible segmentations of heart substructures despite an apparent lack of signal in some non-contrast scans, indicating that it relies on general statistical patterns of heart anatomy.

4.4 Potential Biases in the Study Cohort

Due to the scope of the study, only pseudonymized data were available. Consequently, information such as socioeconomic status, vital status, and ethnicity was not available.

We identified a difference in predictions for males and females, suggesting that the selected features are not equally valid for both sexes. Similar patterns may emerge from other factors.

4.5 Perspectives

The AI model can aid radiologists by quantifying signs of AHF and can assist physicians in early identification of AHF, before a radiology report is available, prompting the need for acute management or further diagnostic procedures. Thresholds should be tested clinically and prospectively, ensuring both sensitivity and specificity exceed high enough impact to support reliable diagnostic decision-making. We consider the model a preliminary step towards automated risk assessment of AHF. Future improvements should focus on developing user-friendliness and validation across external cohorts.

4.6 Study Limitations

We used labels from radiology reports as ground truth and CT scans from a single center consisting of relatively limited amount of training data compared to commercial algorithms, limiting the generaliz-ability. While we had large variation in CT scans and substantial heterogeneity of the cohort, external validation in other cohorts is crucial to assess the robustness and clinical applicability.

Only pseudonymized data and designed features were used, risking population biases, and including only one scan per subject risks excluding the most acute admissions in cases with multiple hospitalizations Lastly, we did not validate in an external cohort or in a prospective manner.

4.7 Conclusion

We developed an explainable AI model that accurately predicts radiological signs of AHF, achieving an AUROC of 0.87, comparable to clinical radiologists. Model interpretability was enhanced through SHAP values for subject-level explanation and supported by visual inspection of segmentations, enhancing model interpretability. This approach supports clinical interpretability of the AI predictions, potentially improving diagnostic speed and accuracy in emergency care setting.

**5. Ethics**

This study ("Diagnostik med kunstig intelligens på CT hos patienter med åndenød: udvikling og validering af maskinlæringsalgoritmer (Breath-CT)") is approved by The Danish Research Ethics Committee (Project-id: 1575037), which also waived informed consent. The study protocol ("Forsøgsprotokol 1.2", approved December 2020) is available from the Danish Research Ethics Committee and the authors.

**6. Availability of software and data**

TotalSegmentator is available from https://github.com/wasserth/TotalSegmentator. The Boosted Trees model is available from https://github.com/dmlc/xgboost. The data is not made publicly available. The trained model is not made publicly available.

**References:**

1. McDonagh TA, Metra M, Adamo M, Gardner RS, Baumbach A, Böhm M, et al. 2021 ESC Guidelines for the diagnosis and treatment of acute and chronic heart failure. Eur Heart J. 2021;42(36):3599–726.

2. Kocher KE, Meurer WJ, Fazel R, Scott PA, Krumholz HM, Nallamothu BK. National trends in use of computed tomography in the emergency department. Ann Emerg Med [Internet]. 2011;58(5):452-462.e3. Available from: http://dx.doi.org/10.1016/j.annemergmed.2011.05.020

3. Yacoub B, Varga-Szemes A, Schoepf UJ, Kabakus IM, Baruah D, Burt JR, et al. Impact of Artificial Intelligence Assistance on Chest CT Interpretation Times: A Prospective Randomized Study. AJR Am J Roentgenol. 2022 Nov;219(5):743–51.

4. Chen T, Guestrin C. Xgboost: A scalable tree boosting system. In: Proceedings of the 22nd acm sigkdd international conference on knowledge discovery and data mining. 2016. p. 785–94.

5. Miger K, Fabricius-Bjerre A, Overgaard Olesen AS, Sajadieh A, Høst N, Køber N, et al. Chest computed tomography features of heart failure: A prospective observational study in patients with acute dyspnea. Cardiol J. 2022;29(2):235–44.

6. Wasserthal J, Meyer M, Breit H-C, Cyriac J, Yang S, Segeroth M. TotalSegmentator: robust segmentation of 104 anatomical structures in CT images. 2022 Aug 11 [cited 2023 Mar 28]; Available from: http://arxiv.org/abs/2208.05868

7. Lundberg SM, Erion G, Chen H, DeGrave A, Prutkin JM, Nair B, et al. From Local Explanations to Global Understanding with Explainable AI for Trees. Nat Mach Intell. 2020 Jan;2(1):56–67.

8. Lanzafame LRM, Bucolo GM, Muscogiuri G, Sironi S, Gaeta M, Ascenti G, et al. Artificial Intelligence in Cardiovascular CT and MR Imaging. Life (Basel, Switzerland). 2023 Feb;13(2).

9. Jain CC, Tschirren J, Reddy YN V, Melenovsky V, Redfield M, Borlaug BA. Subclinical Pulmonary Congestion and Abnormal Hemodynamics in Heart Failure With Preserved Ejection Fraction. JACC Cardiovasc Imaging. 2022 Apr;15(4):629–37.

10. Velichko E, Shariaty F, Orooji M, Pavlov V, Pervunina T, Zavjalov S, et al. Development of computer-aided model to differentiate COVID-19 from pulmonary edema in lung CT scan: EDECOVID-net. Comput Biol Med. 2022 Feb;141:105172.

11. Brady AP, Allen B, Chong J, Kotter E, Kottler N, Mongan J, et al. Developing, Purchasing, Implementing and Monitoring AI Tools in Radiology: Practical Considerations. A Multi-Society Statement From the ACR, CAR, ESR, RANZCR & RSNA. Can Assoc Radiol J [Internet]. 2024; Available from: https://doi.org/10.1186/s13244-023-01541-3

**Tables**

| Dataset | Group | Size | Age | | Outcome |
|---|---|---|---|---|---|
| | | | Range | Mean | Prevalence |
| **Train** | Male | 1597 | 17-99 | 64.89 | 0.076 |
| | Female | 1551 | 18-105 | 68.54 | 0.072 |
| | **Total** | 3148 | 17-105 | 66.69 | 0.074 |
| **Test** | Male | 792 | 18-100 | 64.45 | 0.072 |
| | Female | 732 | 17-100 | 68.24 | 0.093 |
| | **Total** | 1524 | 17-100 | 66.27 | 0.082 |
| **All** | Male | 2359 | 17-100 | 64.75 | 0.075 |
| | Female | 2283 | 17-105 | 68.44 | 0.078 |
| | **Total** | 4672 | 17-105 | 66.55 | 0.077 |

**Table 1** - Data split statistics. The number of males and females is balanced in both training and test data. Females tend to be almost four years older on average. The prevalence is slightly higher for females overall, but slightly higher for males in the training data, and higher for females in the test data.

| Group | Condition | Predicted | | | | TPR FPR P+ | FNR TNR P- | Prevalence |
|---|---|---|---|---|---|---|---|---|
| | | P+ | P- | Total | | | | |
| All | C+ | 63 | 62 | 125 | | 0.50 (0.42-0.59) | 0.50 (0.41-0.58) | 0.08 |
| | C- | 64 | 1335 | 1399 | | 0.05 (0.04-0.06) | 0.95 (0.94-0.96) | |
| | Total | 127 | 1397 | 1524 | **Predicted prevalence** | 0.08 | | |
| Female | C+ | 32 | 36 | 68 | | 0.47 (0.35-0.59) | 0.53 (0.41-0.65) | 0.09 |
| | C- | 21 | 643 | 664 | | 0.03 (0.02-0.05) | 0.97 (0.95-0.98) | |
| | Total | 53 | 679 | 732 | **Predicted prevalence** | 0.07 | | |
| Male | C+ | 31 | 26 | 57 | | 0.54 (0.41-0.67) | 0.46 (0.33-0.59) | 0.07 |
| | C- | 43 | 692 | 735 | | 0.06 (0.04-0.08) | 0.94 (0.92-0.96) | |
| | Total | 74 | 718 | 792 | **Predicted prevalence** | 0.09 | | |

**Table 2 -** Test set classifications using threshold = 0.26840377. P+ = Predicted positive, P- = Predicted negative, C+ = Condition positive (AHF yes), C- = Condition negative (AHF no), TPR = True positive rate, FNR = False negative rate, FPR = False positive rate, TNR = True negative rate. Rates are given with bootstrapped 95% confidence intervals in parentheses.

|  | Very unlikely | Somewhat unlikely | Neutral | Somewhat likely | Very likely | All |
| --- | --- | --- | --- | --- | --- | --- |
| False positives | 22 (34%) | 16 (25%) | 2 (3%) | 13 (20%) | 11 (17%) | 64 |
| False negatives | 11 (18%) | 13 (21%) | 3 (5%) | 14 (23%) | 20 (33%) | 61 |

**Table 3** - Radiologist's review of error cases, showing the radiologist's estimate of likelihood of AHF and pulmonary congestion based on the CT scan, for the false positives and false negatives by the model. A score of "Neutral" was considered when the CT image could not stand alone and there was need for clinical data as AHF could not completely be ruled out.

**Figures**

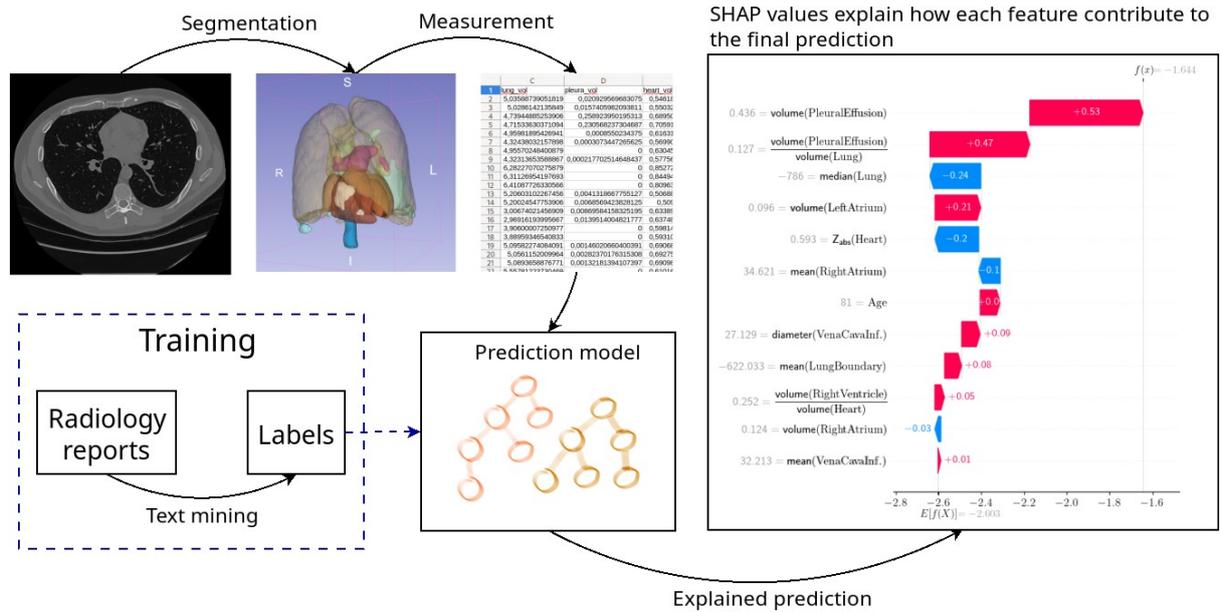

**Figure 1** – Model overview. CT images from PACS were processed with TotalSegmentator to obtain segmentations. Measurements were extracted from the segmentations and passed to the Boosted Trees model. Predictions from the Boosted Trees were passed to SHAP, which provides an explanation of how the features influence the final prediction (log odds ratio of AHF).

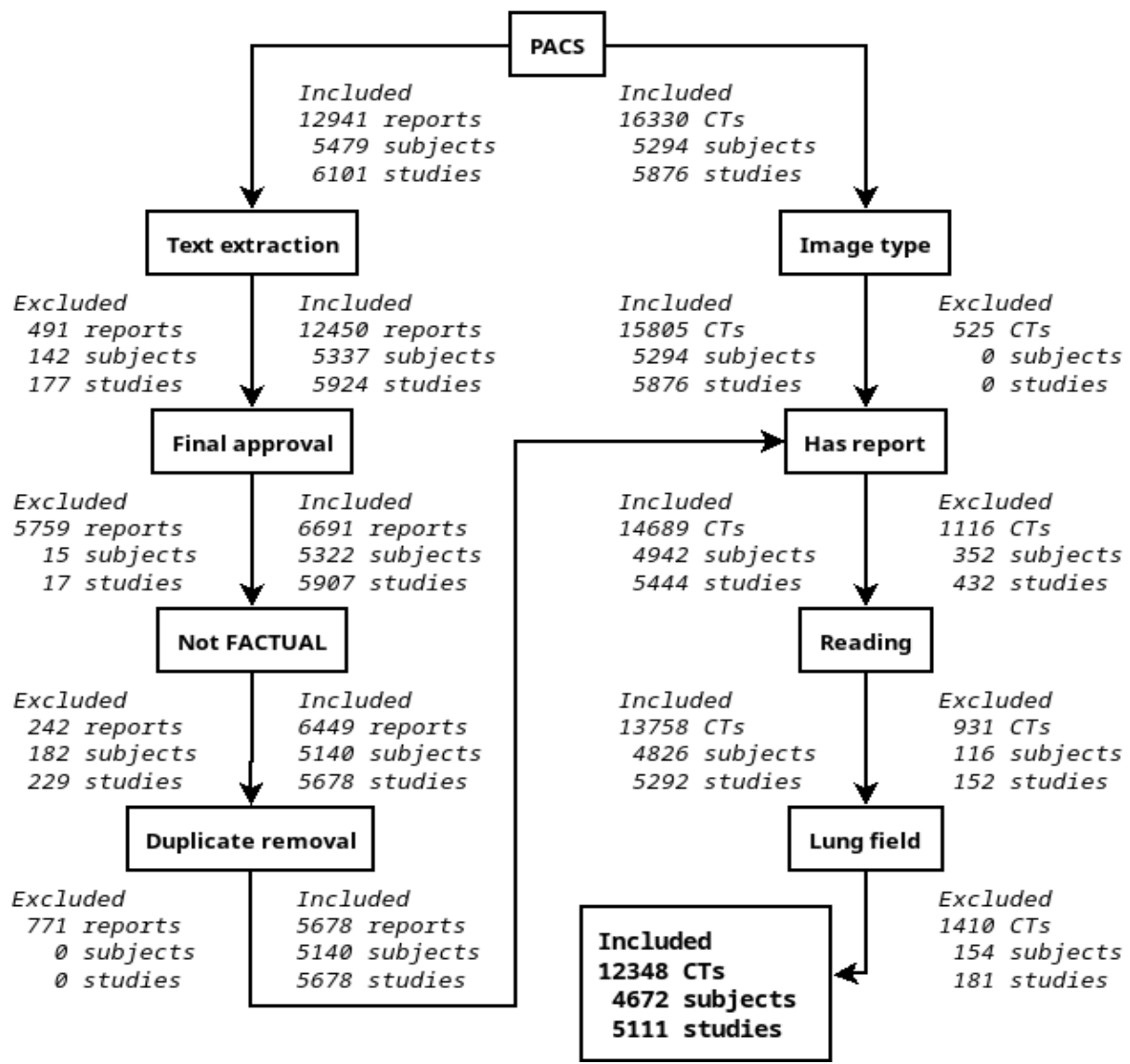

**Figure 2** – Overview of the data flow of inclusion and exclusion, including the exclusion reasons during dataset preparation. Exclusions were applied at multiple stages and included:

"Text extraction": Incomplete/empty report. Formatting errors, "Final approval": Reports not marked as finally approved, "Not FACTUAL": study participants not in the FACTUAL study, "Image type": Image is GE VUE or GE MD, "Has report": No corresponding radiology report available, "Reading": Failed to read DICOM series (e.g. corrupted file, missing series instances, no image data), "Lung field": Non thoracic-scans (e.g. incorrectly stored head/neck scan) or part of lung field not visible (e.g. cropped to heart).

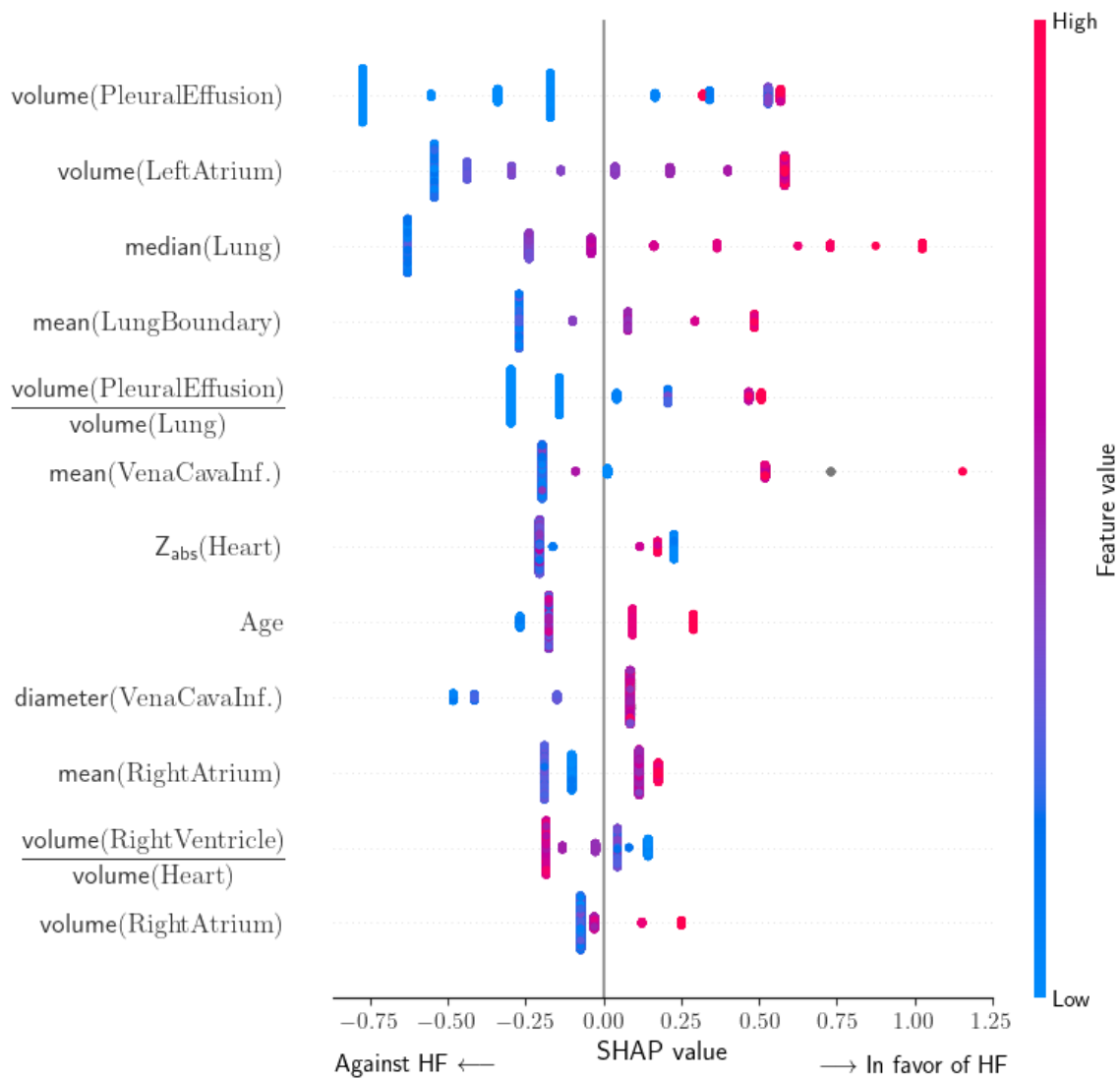

**Figure 3** – A Beeswarm plot of SHAP values for the final selected features. Each dot represents a scan with a valid report label. A positive SHAP value means that the feature for that scan increases the likelihood of a positive prediction. The coloring indicates the relative feature value. e.g., light blue for volume (Pleural Effusion) means zero pleural effusion and light blue for median (Lung) means large negative HU value. Gray dots, like the one for mean (VCI), with a SHAP value around 0.75, indicate missing values. In this case due to a bad segmentation of the vena cava inferior. *HF: Heart Failure.*

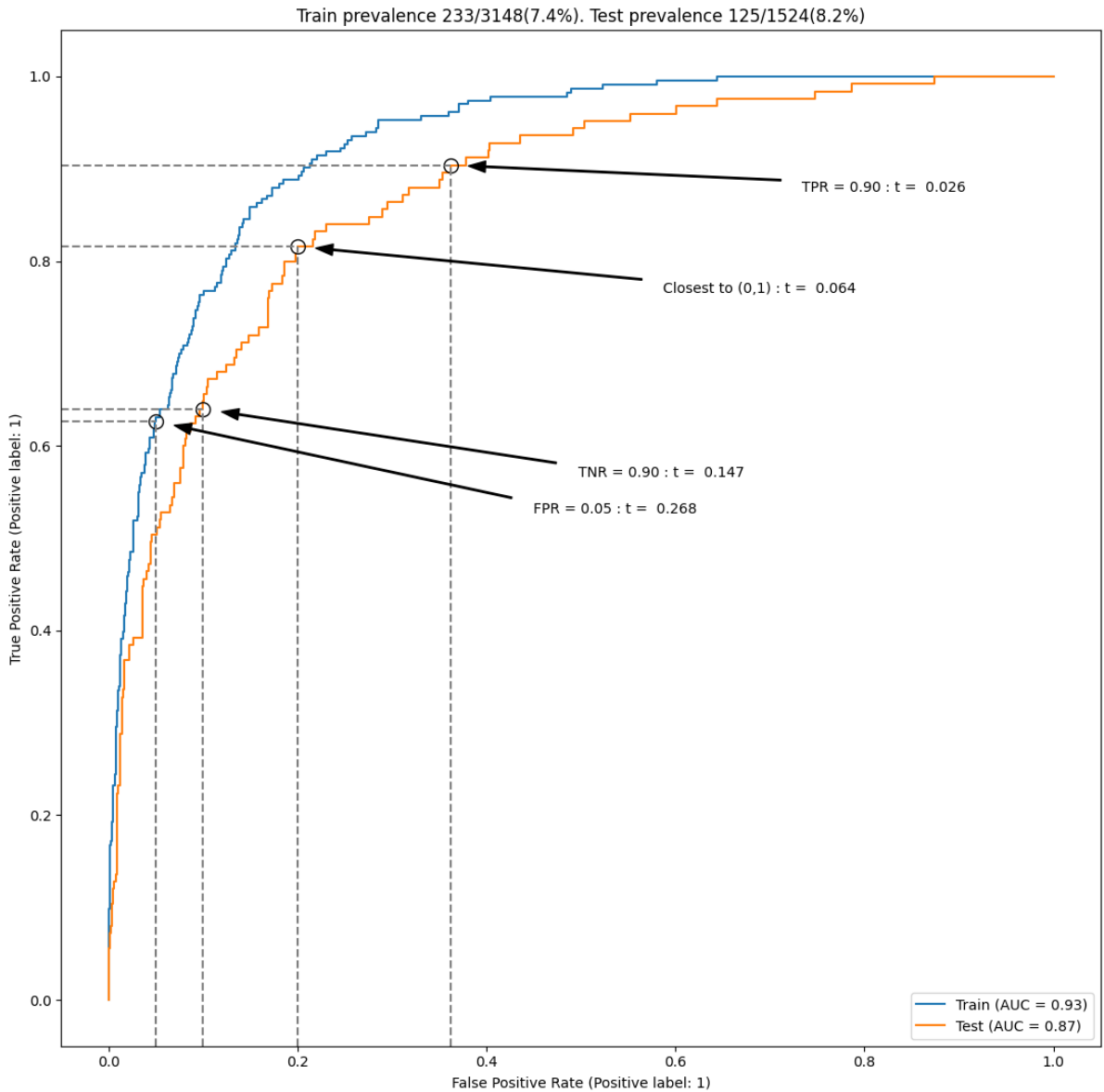

**Figure 4 -** ROC curve for predicting radiological signs of AHF on train data (blue line, AUROC=0.93) and test data (orange line, AUROC=0.87). The four circles indicate points of interest on the ROC curve. From left to right, 0.05 false-positive-rate (FPR) on train data (t=0.268), used to determine decision threshold for test data; 0.90 true-negative-rate (TNR) on test data (t=0.148); closest point to "perfect" on test data (t=0.064); 0.90 true-positive-rate (TPR) on test data (t=0.028).

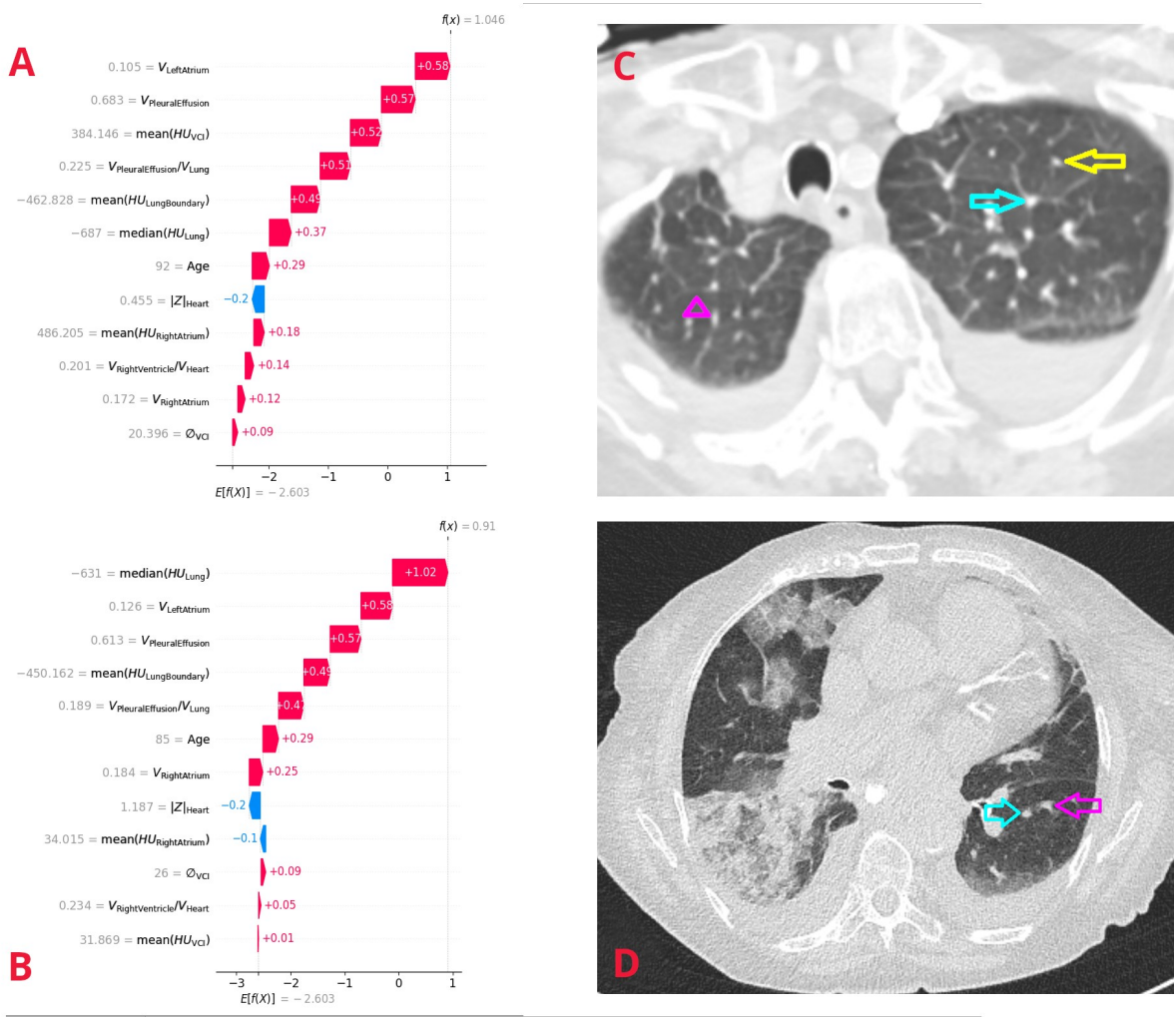

**Figure 5 -** SHAP waterfall plots illustrating two false positive cases (A and B). Red arrows (pointing to the right) indicate features contributing to a positive prediction, and blue errors (pointing to the left) indicate features contributing to a negative prediction. Expert review considered patient A as "very probable" to have AHF, suggesting a true positive prediction from the model and a mislabeling from the radiology report. Subfigure C shows examples of radiological findings for case A: smooth septal thickening (magenta arrowhead) and a dilated vein (teal arrow) larger than the artery (yellow arrow) as well as bilateral pleural effusion. Further details are in Supplemental Figure S6. Patient B was considered "very unlikely" to have AHF and was therefore assessed as false positive. Subfigure D shows some of the findings for case B: bilateral pleural effusion but without dilation of vessels and interstitial

oedema. The arrows mark the vein (teal arrow) and artery (magenta arrow). Further details are in Supplemental figure S5. The mean log odds, estimated on the train data, is indicated at the bottom of the plot with [f(x)] = -2.603, corresponding to a predicted prevalence of approximately 7%.

**Supplemental Material**

| Finding | Regular expression (Danish) |
|---|---|
| Congestion | \b(lunge)?stase |
| No congestion | \b((ingen)|(uden(?! kontrast))|(ikke(?! overbevisende)))[ \w,]*\b(lunge)?stase |
| Heart failure | hjertesvigt |
| No heart failure | \b((ingen)|(uden(?! kontrast))|(ikke))[ \w,]*\bhjertesvigt |
| Decompensation | (in[ck]ompensation)|(hjerteinsufficiens) |
| No decompensation | \b((ingen)|(uden(?! kontrast))|(ikke))[ \w,]*\b((in[ck]ompensation)|(hjerteinsufficiens)) |
| Edema | \blunge[ ]*ødem |
| No edema | \bingen[ \w]*lunge[ ]*ødem |
| Pleural effusion | pleura(le?)?[ ]*((væske)|(ansamling)|(effusion)) |
| No pleural effusion | \b((ingen)|(uden(?! kontrast))|(ikke))[ \w,]*(perikardie-)?[ \w,]*\bpleura(le?)?(-[ \w]*)?[ ]*((væske)|(ansamling)|(effusion)) |

**Table S1** - Regular expressions used for mining the reports.

| Parameter | Space | Initial | Final |
|---|---|---|---|
| eta | 0.05, 0.1, 0.2, 0.3 | 0.2 | 0.2 |
| gamma | 0, 1 | 0 | 0 |
| max_depth | 1, 2, 3 | 2 | 1 |
| min_child_weight | 0, 1 | 1 | 0 |
| max_delta_step | 0 | 0 | 0 |
| subsample | 0.5, 1.0 | 0.5 | 1.0 |
| lambda | 0, 1 | 0 | 0 |
| alpha | 0, 1, 2, 3, 4, 8 | 3 | 4 |
| tree_method | auto | auto | auto |
| scale_pos_weight | 1, balanced | 1 | 1 |

**Table S2** - Parameter search space, with initial and final selected parameters. The definitions of the parameters can be found in (4).

| | | |
|---|---|---|
| Manufacturer | | |
| | Phillips | 1251 |
| | SIEMENS | 10234 |
| | GE | 711 |
| | TOSHIBA | 152 |
| Image type | | |
| | Single energy | 5371 |
| | Dual energy | 6977 |
| Reconstruction kernel | | |
| | Smooth | 1743 |
| | Medium | 5190 |
| | Sharp | 5415 |
| | Number of different kernels | 50 |
| Contrast | | |
| | With contrast | 10191 |
| | Without contrast | 2157 |

**Table S3** – CT-information

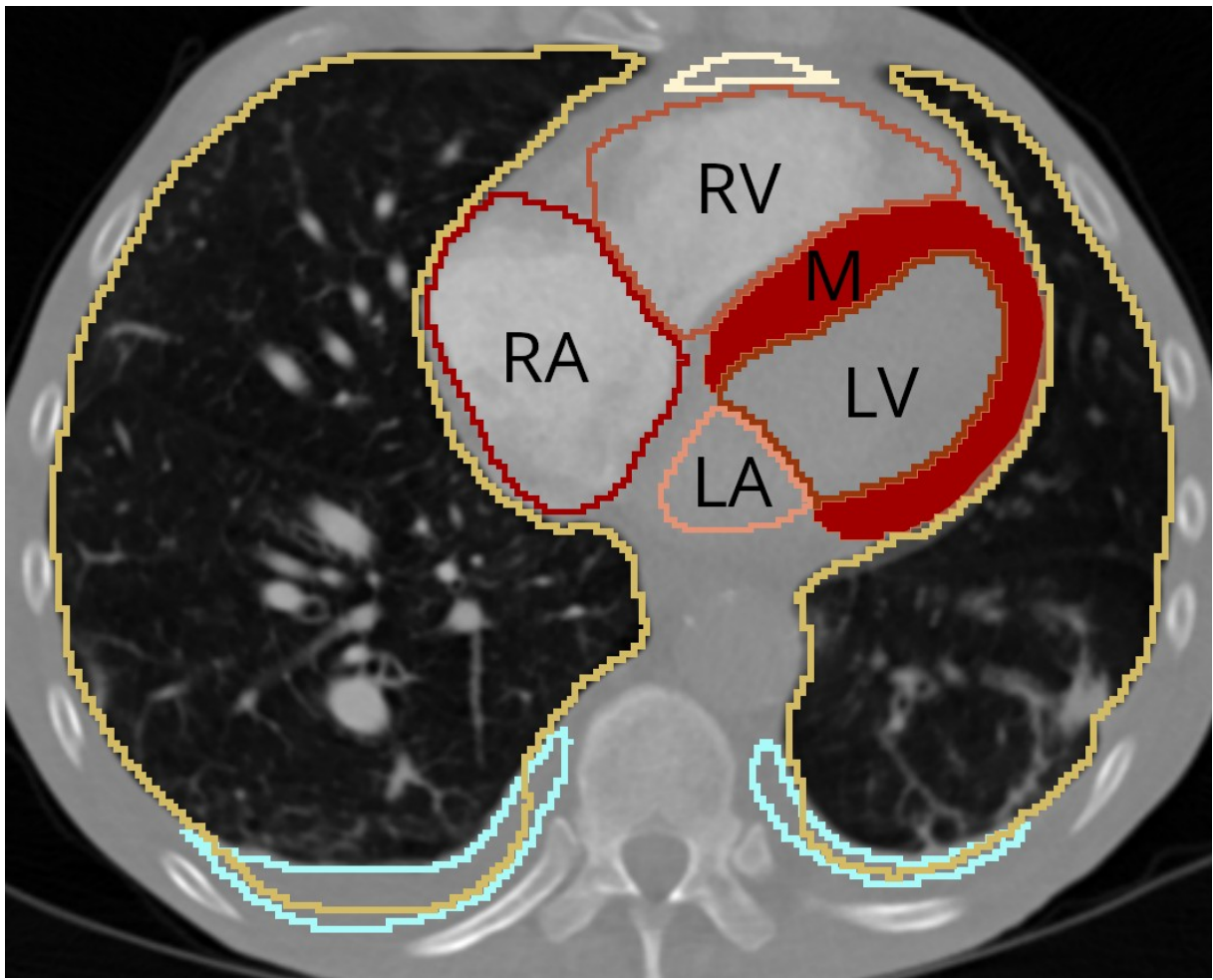

**Figure S1** – A random example of segmentations from TotalSegmentator. The heart is defined as the union of left atrium (LA), right atrium (RA), left ventricle (LV), right ventricle (RV) and myocardium (M). Heart segmentations were acceptable, though minor discrepancies – such as underestimation of LA area as seen here – may reflect either segmentation limitations or variations in slice level.

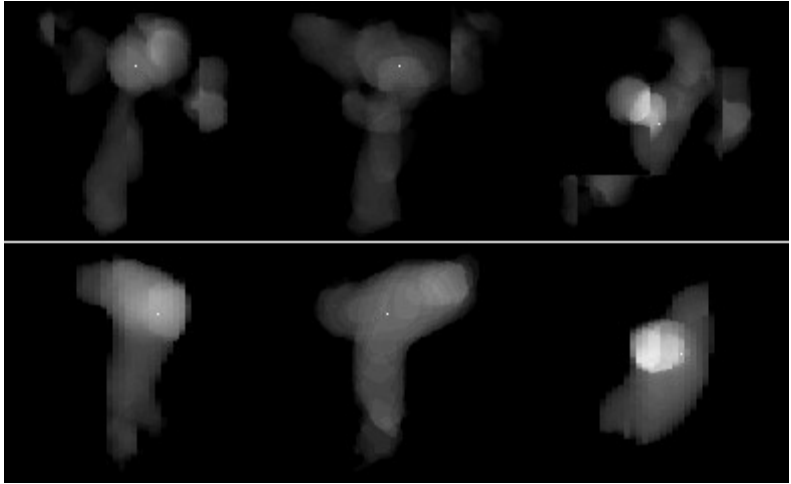

**Figure S2** - Axis parallel projections of pulmonary artery segmentations, used to assess whether the segmentations appear to have the correct shape. Segmentations where the artery shapes are anatomically plausible are deemed to be of reasonable quality. The top segmentation is of poor quality, the bottom is of reasonable quality. Lung field segmentations were assessed similarly.

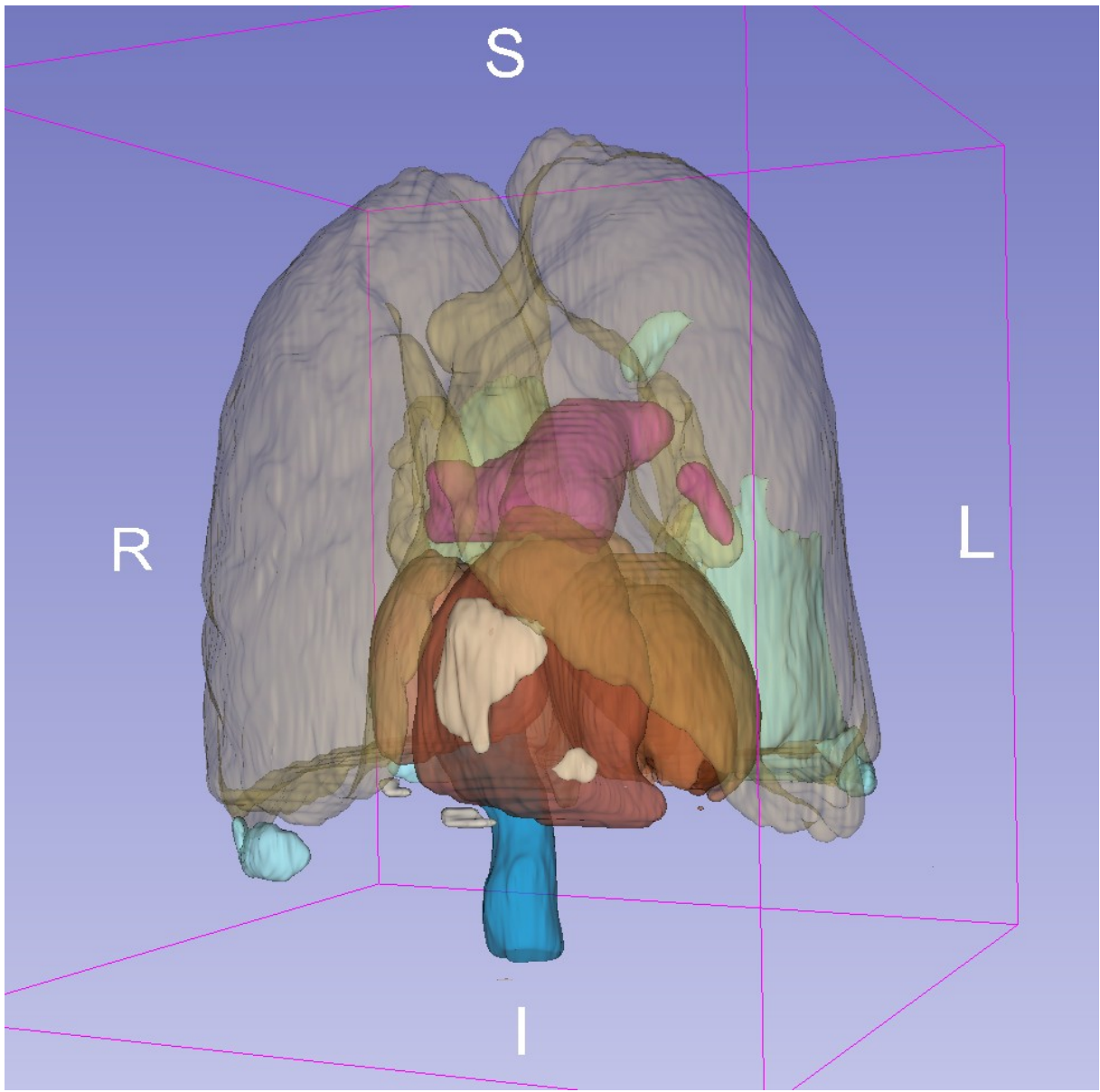

**Figure S3** - Randomly selected example of the segmentation. Complete 3D visualizations of selected segmented CT volumes were visually assessed to check correctness of the segmentations.

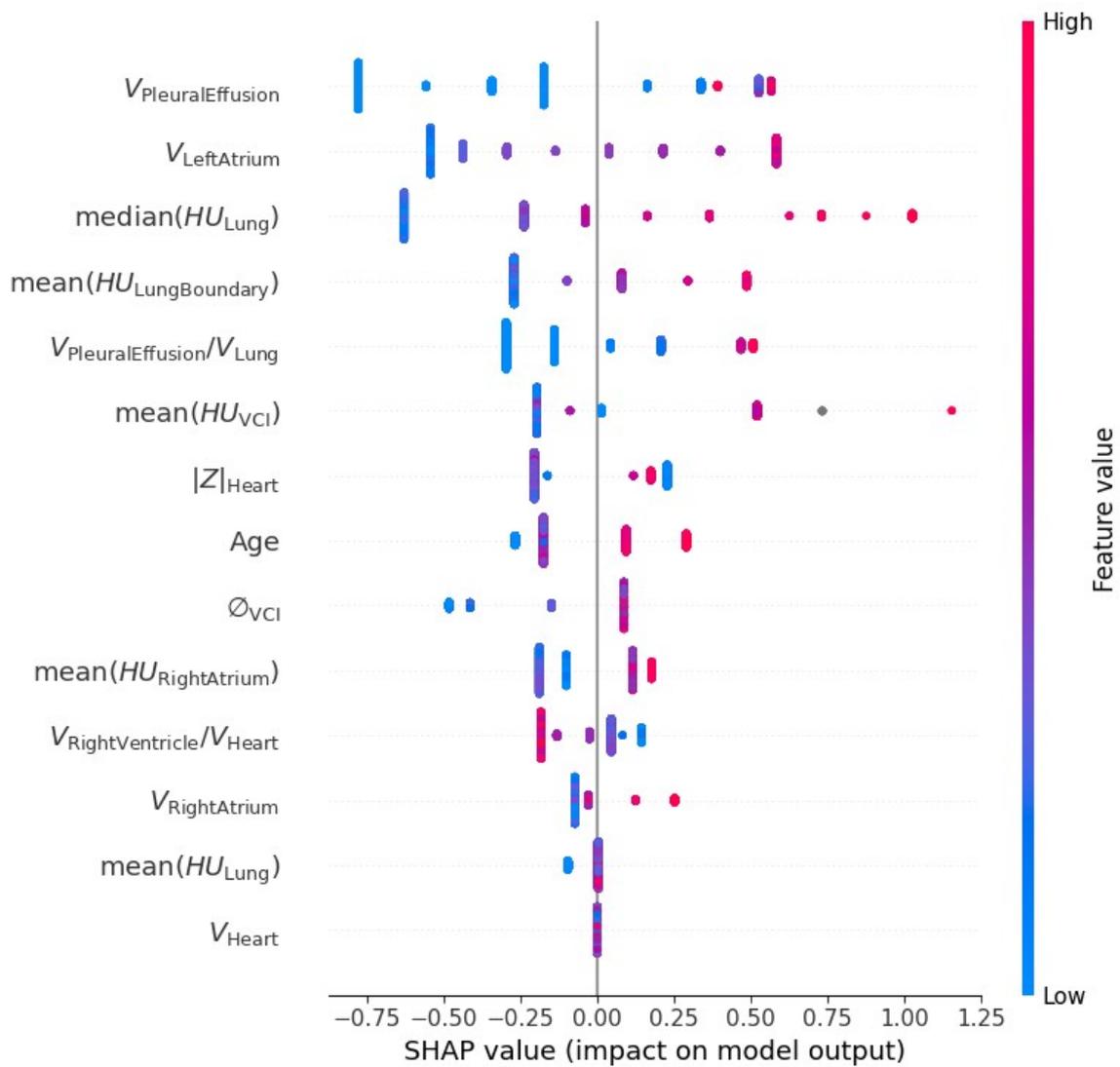

**Figure S4** - Beeswarm plot of feature importance, as determined by SHAP values, prior to manual pruning.

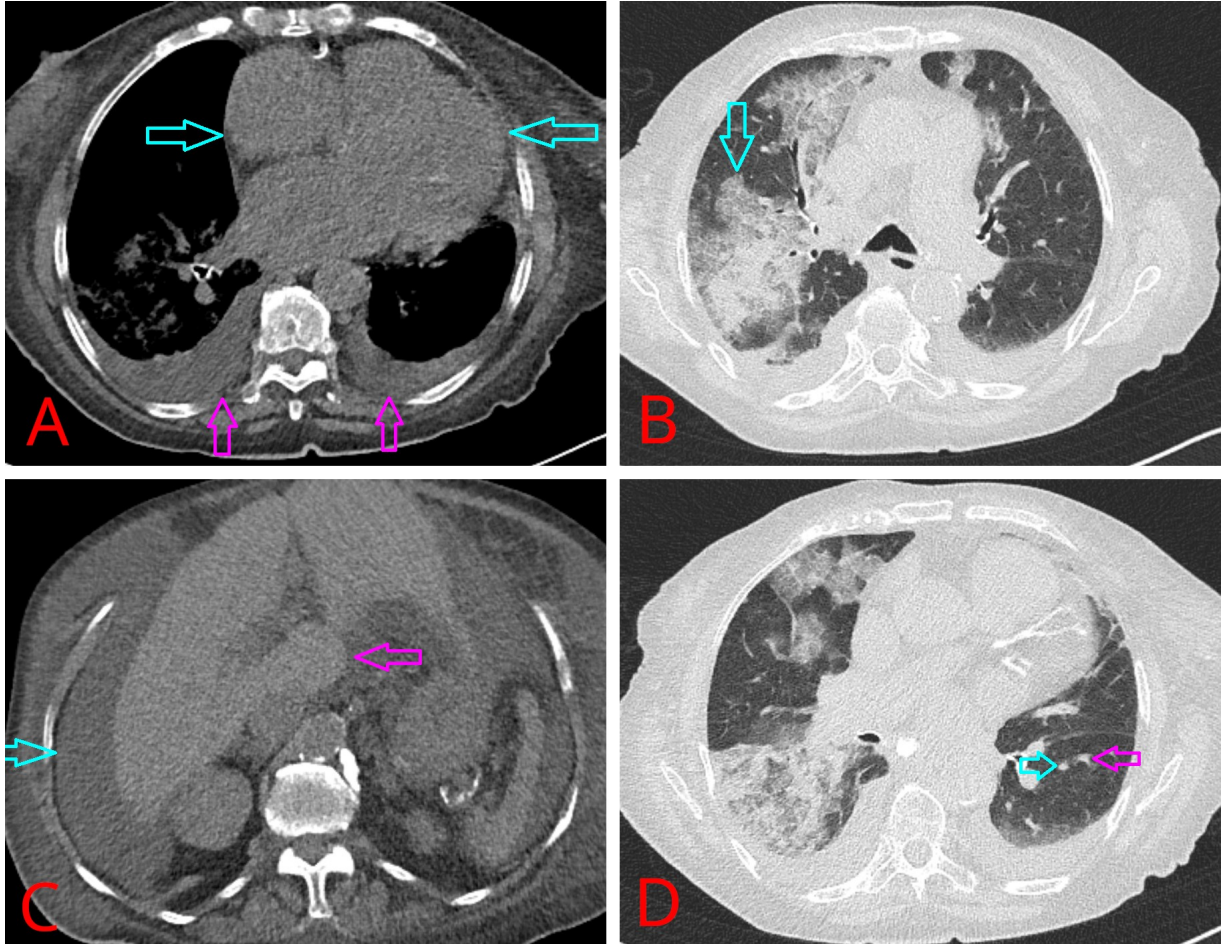

**Figure S5** - Error review of false positive reviewed as" Very unlikely" for AHF. A: Enlarged heart (teal arrow) but without interstitial oedema. Parenchymal bleeding and cirrhosis (magenta arrows). B: Infiltrations (teal arrow) with probable parenchymal bleeding. C: Acites caused by liver cirrhosis (teal arrow) and pleural effusion. Hypertrophy (magenta arrow). D: Bilateral pleural effusion without dilation of vessels and without interstitial oedema. The vessels are marked as vein (teal arrow) and an artery (magenta arrow).

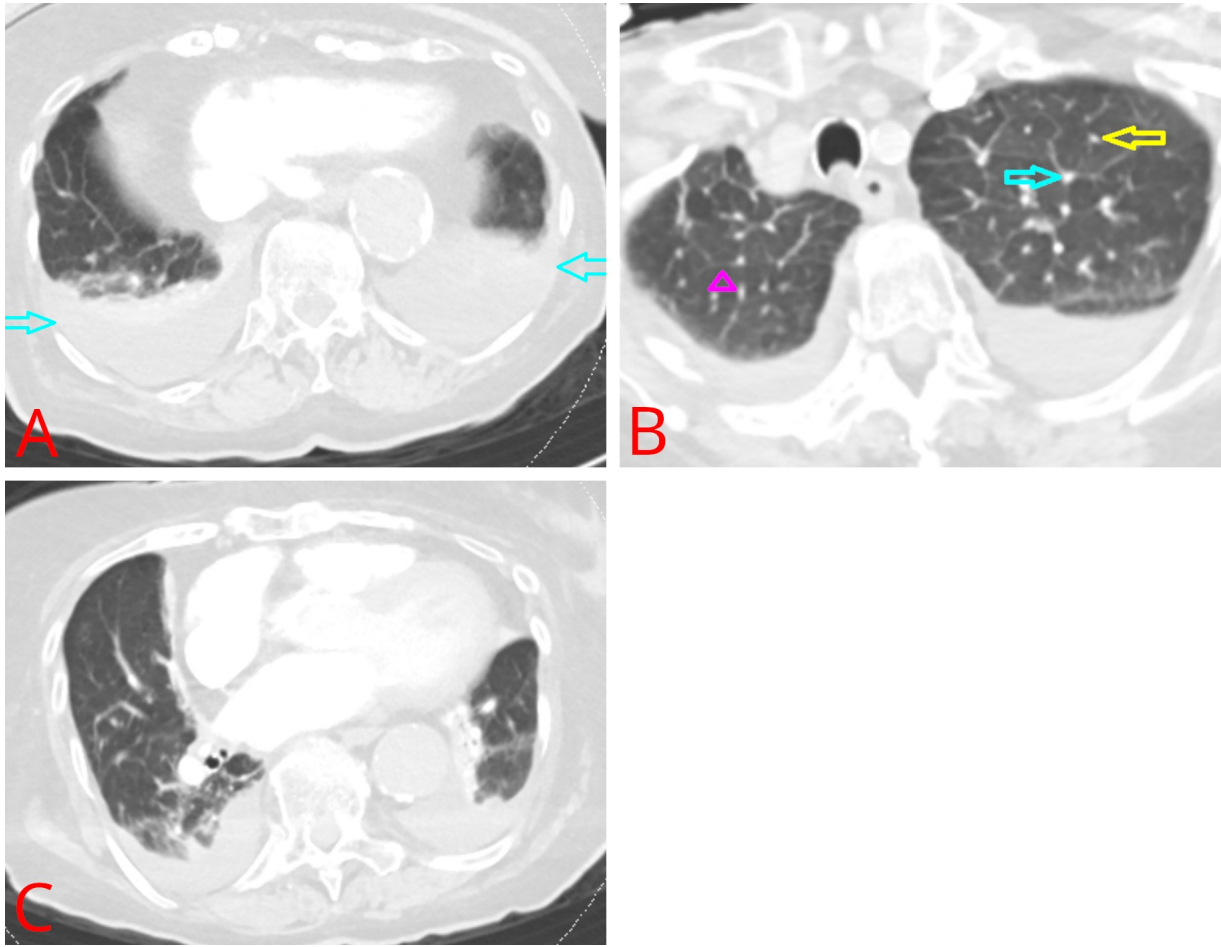

**Figure S6** - Error review of false positives reviewed as "Very probable" for AHF. A: Bilateral pleural effusion (teal arrows). B: Septal thickening (magenta arrowhead) along with dilated vessels, where the vein (teal arrow) is larger than the artery (yellow arrow). C: Slightly enlarged heart and bilateral pleural effusion.